Correspondence and requests for materials should be addressed to M.S.-P. (marta.sales@urv.cat)

# Predicting future conflict between team-members with parameter-free models of social networks

Núria Rovira-Asenjo[1], Tània Gumí[1], Marta Sales-Pardo[1] & Roger Guimerà[1,2]

[1]Departament d'Enginyeria Química, Universitat Rovira i Virgili, Av. Països Catalans 26, Tarragona 43007, Catalonia, Spain, [2]Institució Catalana de Recerca i Estudis Avançats (ICREA), Lluís Companys 23, Barcelona 08010, Catalonia, Spain.

**Despite the well-documented benefits of working in teams, teamwork also results in communication, coordination and management costs, and may lead to personal conflict between team members. In a context where teams play an increasingly important role, it is of major importance to understand conflict and to develop diagnostic tools to avert it. Here, we investigate empirically whether it is possible to quantitatively predict future conflict in small teams using parameter-free models of social network structure. We analyze data of conflict appearance and resolution between 86 team members in 16 small teams, all working in a real project for nine consecutive months. We find that group-based models of complex networks successfully anticipate conflict in small teams whereas micro-based models of structural balance, which have been traditionally used to model conflict, do not.**

Teamwork is increasingly important. In science, where teamwork has been best studied quantitatively thanks to the large amount of data available, studies have forecast the shift from individual work to teams for over a century[1]. Indeed, the classic studies of de Solla Price predicted that by 1980 no articles in chemistry would be authored by single authors[2]. Although these predictions have not come true, recent studies indicate that teamwork has become more frequent in virtually all fields and subfields of science[3,4]. Parallel to this, there has been an increase in the impact of works produced by teams to the point that, today, the most highly cited works in all fields are overwhelmingly produced by teams[4,5]. This tendency is due in part to external factors such as the increased complexity of cutting edge research[3], the widespread use of new technologies[5], the growth of the number of researchers, and the trend towards greater specialization[4]. There is also mounting evidence that diversity provides an intrinsic advantage to teams, and that teams composed by diverse individuals have higher performance than teams composed by similar individuals[6,7,3]. All in all, the "collective intelligence factor" of a team is a better predictor of team performance than the abilities of each team member[8].

Despite the benefits of teams, teamwork also results in communication, coordination and management costs[9,10]. More importantly, conflict arises in teams from tension among members. Conflict (in particular, conflict that is related to personal relationships) is known to interfere with team functioning and may offset the benefits of teamwork[6,11–14]. In a context where teams play an increasingly important role, it is important to understand conflict and to develop diagnostic tools to avert it.

Here, we investigate empirically whether it is possible to quantitatively predict future conflict in small teams. Rather than using regression analysis for "conflict forensics" (that is, to explain a posteriori what factors correlate with higher levels of conflict in a given team)[13,15–17], we focus on first-principles parameter-free models of social network structure, and on prediction rather than postdiction. As recently suggested[18,19], analyzing teams as networks poses the methodological dilemma of choosing between "micro-level" sociopsychological theories such as structural balance[20–24] and "macro-level" theories developed in the context of network science[25–27,19]. We show that, paradoxically, statistical network methods can successfully anticipate conflict in small teams whereas some of the most widely-used micro-level sociological theories cannot.

## Results

**Measuring conflict and conflict evolution in small teams.** Our study draws upon a long history of network experiments with teams and small groups, dating back to the experiments carried out in the 50's by the Group Networks Laboratory at MIT[28,19]. We analyze data from 16 small teams with 3 to 7 members each, for a total of 86 team members and 374 reported within-team interactions between them (Table 1). All teams worked in the same open-ended project for nine consecutive months. Importantly, the teams we analyze were facing a real task as opposed to a simplified experimental task, and their members had real incentives and experienced real conflicts that developed throughout the extended duration of the project.

Table 1 | Teams and reported interactions between team members

| Team | Team size | Number of interactions *YY* | *YN* | *NY* | *NN* |
|---|---|---|---|---|---|
| A | 5 | 10 | 4 | 3 | 3 |
| B | 6 | 20 | 3 | 2 | 5 |
| C | 6 | 26 | 0 | 3 | 1 |
| D | 5 | 12 | 3 | 1 | 4 |
| F | 5 | 10 | 5 | 0 | 5 |
| G | 3 | 5 | 0 | 1 | 0 |
| H | 4 | 7 | 4 | 0 | 1 |
| I | 3 | 3 | 0 | 1 | 2 |
| J | 7 | 29 | 7 | 1 | 5 |
| K | 6 | 24 | 2 | 0 | 4 |
| L | 7 | 18 | 3 | 10 | 11 |
| M | 6 | 21 | 0 | 4 | 5 |
| N | 5 | 16 | 1 | 2 | 1 |
| O | 7 | 34 | 3 | 1 | 4 |
| P | 4 | 12 | 0 | 0 | 0 |
| Q | 4 | 10 | 1 | 1 | 0 |
| **Total** | 83 | 257 | 36 | 30 | 51 |

To track the development of conflict in the teams, we administered the same survey to all team members twice, at the middle and end of the project (surveys I and II, four and nine months into the project, respectively). In the survey we asked all individuals about their disposition to work with each of the other team members in the future. We use the answers to this question to construct two directed networks for each team in which a link from member *A* to member *B* can be of two types, namely, $l_{AB} = Y$ if the answer was positive (*A* is willing to work with *B* in the future) and $l_{AB} = N$ if the answer was negative (Fig. 1). We use changes in link type between the two surveys as a proxy for conflict appearance/resolution: conflict appears when a link of type *Y* in survey I becomes *N* in survey II; conflict resolves when a link of type *N* in survey I becomes *Y* in survey II. In what follows, we will denote interactions between team members as $l^I l^{II}$, $l^I$ being the link type from survey I and $l^{II}$ the link type from survey II (for example *YN* denotes a link where conflict arose during the project). Of the 374 reported interactions between individuals, 257 were *YY*, 36 were *YN*, 30 were *NY* and 51 were *NN* (Table 1).

**Structural balance versus block model-based link reliability.** In social network analysis, conflict evolution has traditionally been studied using the concept of "balance," which focuses on the state of network triads (or, more generally, network cycles)[20–24]. In a directed graph, a triad is in a balanced state when there is an odd number of positive reciprocal connections between individuals[22]; otherwise, a triad is in an unbalanced state. According to this theory, unbalanced states produce tension and generate changes towards balance[20]. For example, if *A* and *B* have a positive relationship and so do *A* and *C*, then if *B* and *C* have a negative relationship (so that the number of positive reciprocal interactions in the triad is two) there is a tension pushing towards either the *B* – *C* relationship becoming positive or one of the others becoming negative. Since the idea of balance revolves around the relationships between small groups of individuals (in this sense we say that it is a "micro-level" theory), it seems a priori well-suited to study the evolution of conflicts in teams.

At the other end of the spectrum of social network models, block models postulate that social actors can be classified into groups such that all actors within a group have similar patterns of interactions with actors in other groups[29–32]. These are "macro-level" models in which the fundamental unit of the models is the group, not the individual, and therefore seem a priori less well-suited to study small teams. However, methods based on block model inference are known to accurately identify reliable and unreliable interactions in large complex networks[32].

Given these considerations, we compare the ability of structural balance theory to predict conflicts within teams to that of a statistical method that uses block models to describe team interactions (Fig. 1).

In particular, we are interested in predicting the state of each link $l^{II}_{AB}$ in the second survey, based on the structure of the team network in the first survey using two methods: the structural balance method (SB) and the link reliability method (LR). SB focuses on the balance of relations induced by the presence of a positive (*Y*) interaction from member *A* to member *B*. In particular, we define the $S^{SB}$ score of each link as the difference between the number of balanced triangles $t_{\mathrm{bal}}$ within the team when $l^I_{AB} = Y$ and the number of balanced triangles within the team when $l^I_{AB} = N$, that is $S^{SB}_{AB} = t_{\mathrm{bal}}(Y) - t_{\mathrm{bal}}(N)$ (Fig. 1c).

In contrast, LR uses a Bayesian approach to sample over all possible stochastic block models of a network to estimate the "reliability" $S^{LR}$ of each link, that is, the probability that the link is of type *Y* based on the observation of the whole team network obtained from survey I (Fig. 1b and Methods)[32,33].

**Conflict prediction performance.** Note that whereas the LR method assigns a probability for each link to become *Y* or *N*, the SB method does not, thus we cannot directly compare outputs from the two methods for each of the links. To compare both methods we analyze instead their ability to rank links within teams. From a ranking perspective, we expect that the higher the score the larger the probability that the link is of type *Y* in survey II; conversely, the lower the score, the larger the probability that a link is of type *N* in survey II.

To measure the ranking accuracy in the case of conflict appearance, we take, for each team, all possible (*YY*, *YN*) link pairs and calculate the number of times the *YY* link in the pair has a higher score than the *YN* link in the pair, according to each method. Conversely, for conflict resolution, we record the number of times that the *NY* link has a score higher than the *NN* for all possible (*NY*, *NN*) link pairs within each team.

For the LR method we find that *YY* links have higher scores than *YN* links 61% of the time (conflict appearance), and that *NY* links have scores higher than *NN* links 67% of the time (conflict resolution). This means that, using the LR method, links with a lower score are consistently more likely to produce conflict in the future (survey II), both when conflict exists and when it does not exist at the time of survey I. In contrast, for the SB method, *YY* links have higher scores than *YN* links only in 47% of the cases, and *NY* links have scores higher than *NN* links in 55% of the cases.

To assess the significance of these results we proceed as follows. For conflict appearance, we consider the ratio $n_{YY}/n_{YN}$ between the number of times that the score of a *YY* link is higher than the score of a *YN* link ($n_{YY}$) and the number of times the reverse is true ($n_{YN}$). Analogously, for conflict resolution we consider the ratio $n_{NY}/n_{NN}$ between the number of times that the score of a *NY* link is higher than the score of a *NN* link ($n_{NY}$) and the number of times the the reverse is true ($n_{NN}$). We denote these ratios as the normalized prediction performance for the appearance of conflict ($n_{YY}/n_{YN}$) and for the resolution of conflict ($n_{NY}/n_{NN}$) (Fig. 2a), respectively. We compare the values obtained for the SB and LR methods to those obtained by resampling the scores of all links, which corresponds to a null model in which links are not separated at all. We find that, at a 5% significance level, the LR method is significantly more accurate than the null model at predicting both the appearance (with p-value $p = 0.030$) and resolution ($p = 0.032$) of conflicts. In contrast, the SB method is not ($p = 0.704$ and $p = 0.232$, respectively).

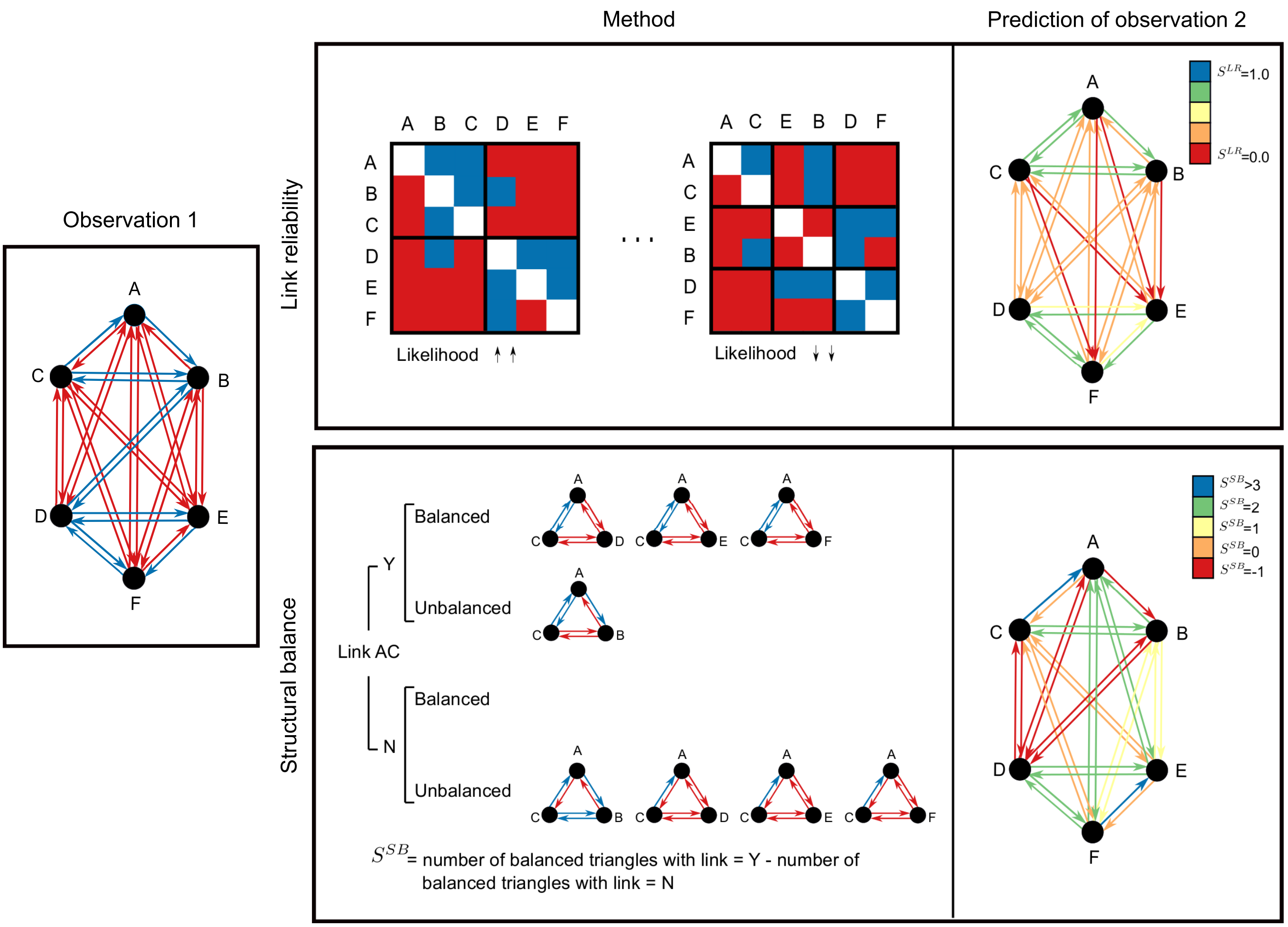

**Figure 1 | Parameter-free network methods for conflict prediction.** (a) For each team, we build a network using the information from survey I (Methods). A blue link from A to B means that A would like to work with B in the future so that $l^{I}_{AB} = Y$. A red link from B to A means the opposite so that $l^{I}_{BA} = N$. To predict which links are more likely to be Y (or N) in survey II (Methods), we apply two different methods: link reliability (LR) (b) and structural balance (SB) (c). (b) The LR method samples all possible partitions of nodes into groups. For each partition, it calculates the probability that $l^{II}_{AB} = Y$ according to that partition. The total probability that $l^{II}_{AB} = Y$ (reliability) is then a weighted sum of these probabilities over all possible partitions (Methods). The weight (likelihood) of a partition depends on how well it describes network connectivity. As an illustration, we show the matrix representation of two partitions. Each row/column corresponds to a node. Matrix elements show link types Y or N color coded in blue and red, respectively. The matrix on the left has a high likelihood because nodes in the same group have similar connection patterns; the matrix on the right has a low likelihood because nodes in the same group have different connection patterns. Finally, we use the reliability scores for each connection to obtain a prediction for observation 2. Link reliability values are color coded following the color bar. (c) The SB theory assumes that a balanced triad exists when there is an odd number of reciprocal relations. To obtain a score $S^{SB}$ for every link, we count the number of balanced triangles in the network $t_{bal}$ when $l^{I} = Y$ minus the number of balanced triangles in the network when $l^{I} = N$. Note that $S^{SB}$ only depends on triangles that include the link of interest. For instance, when $l^{I}_{AC} = Y$, there are three balanced triangles involving $l_{AC}$, while when $l^{I}_{AC} = N$, there are no balanced triangles that involve $l_{AC}$ thus $S^{SB}_{AC} = 3$. We use these scores to build a prediction for observation 2. Link scores are color coded following the color bar.

**Overlap between methods and hybrid scores.** Although the SB method does not seem to consistently predict neither future conflict resolution nor appearance, it may still be possible that it captures different information from that captured by the LR method, so that the predictions of both methods are complementary (Fig. 2b). For conflict appearance, we find that the LR method accurately ranks $n_{YY} = 396$ ($YY$, $YN$) link pairs, whereas the SB method accurately ranks $n_{YY} = 305$ pairs, of which 221 pairs match up in both methods. For conflict resolution, we find that the LR method accurately ranks $n_{NY} = 114$ ($NY$, $NN$) link pairs, whereas the SB method accurately ranks $n_{NY} = 93$, of which 70 pairs match up in both methods.

Since the predictions of the SB method are not a perfect subset of the predictions of the LR method, it is interesting to see if a simple combination of both methods can provide a better prediction of conflict evolution than each of the two methods separately. To investigate this, we define a hybrid score $S^{H}$ that linearly combines the scores of both methods, $S^{H} = \alpha S^{LR} + (1-\alpha)\tilde{S}^{SB}$, where $\tilde{S}^{SB}$ is a properly normalized version of $S^{SB}$ and $\alpha$ is a parameter that enables us to interpolate between each one of the original methods (Methods). As we show in Fig. 3 this hybrid score does not improve, in general, the predictions of the LR method. For conflict appearance, even a small contribution of the SB score is enough to offset the predictive power of the LR method. That is not the case for conflict resolution, but in any case predictions do not significantly improve those of the pure LR method.

## Discussion

Our contributions are of methodological and practical importance for team science. While conflict has long been recognized as one of the main issues in team performance, it is very hard to predict in small teams, precisely because the small size of the teams leaves us with little information about what factors are truly driving conflict dynamics. This poses a methodological challenge that we have addressed by investigating whether micro-based models of structural balance or macro group-based models are more appropriate to tackle the problem. Our results demonstrate that it is possible (albeit difficult) to predict conflict in small teams. Specifically, we find that group-based models have more predictive power, which suggests that the lack of data is better addressed by the complete probabilistic treatment that these models make possible, than by the more detailed models of team dynamics. The immediate practical implication of this finding is that, to avert conflict, groups can in principle be monitored in non-invasive ways (since only the network structure is needed, as opposed to, for example, detailed psychological accounts of team members). Our results thus highlight the relevance of the agenda put forward by Katz and coworkers, when they called for bringing the network perspective back into team science[18].

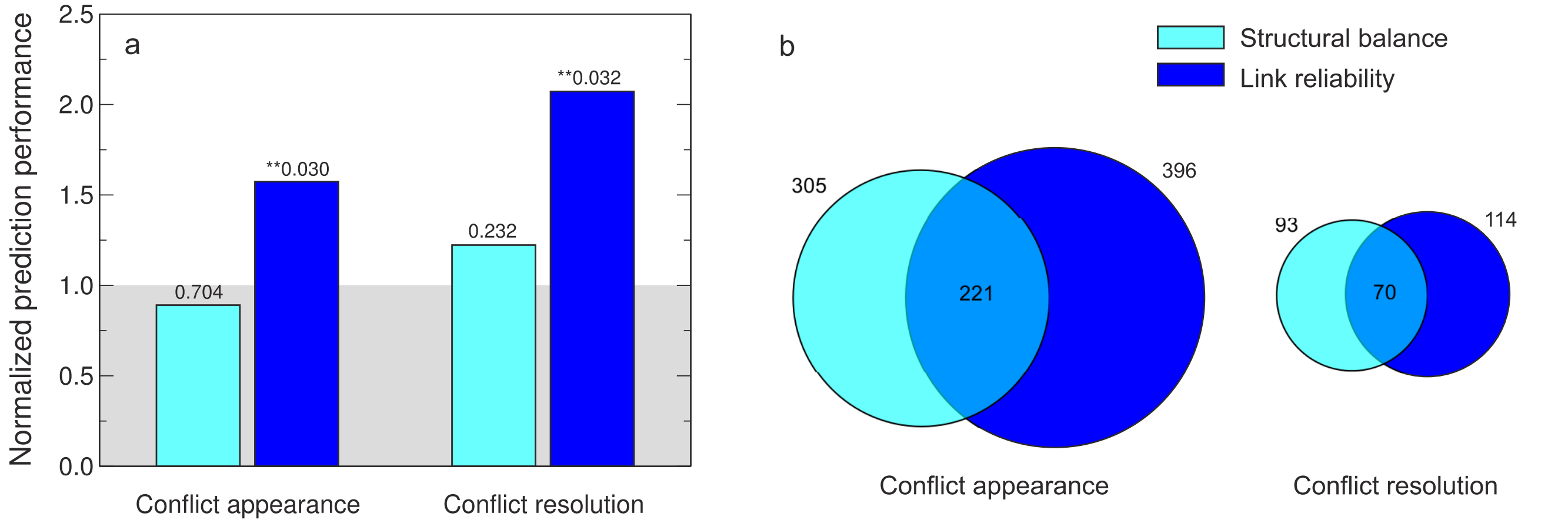

**Figure 2 | Performance of parameter-free network methods for conflict prediction.** (a) We show the performance of the LR (blue) and the SB (cyan) methods, for conflict appearance and resolution. For conflict appearance we consider the ratio between the number $n_{YY}$ of times that the score of a *YY* link (positive in surveys I and II) is higher than the score of a *YN* link (positive in survey I and negative in survey II) in the same team, and the number $n_{YN}$ of times the reverse is true. Analogously, for conflict resolution we consider the ratio between the number $n_{NY}$ of times that the score of a *NY* link is higher than the score of a *NN* link, and the number $n_{NN}$ of times the the reverse is true. We denote these ratios as the normalized prediction performance for the appearance of conflict ($n_{YY}/n_{YN}$) and for the resolution of conflict ($n_{NY}/n_{NN}$). To establish the significance of these results, we compare the values of the normalized prediction performance obtained for the SB and LR methods to those of the null model obtained by resampling the scores of all links within each team. We find that the LR method is significantly more accurate than the null model ($p = 0.030$ for conflict appearance and $p = 0.032$ for conflict resolution), whereas the SB method is not ($p = 0.704$ for conflict appearance and $p = 0.232$ for conflict resolution). (b) We show the overlap of LR and SB methods, for conflict appearance and resolution. The numbers in the figure indicate the number of correctly ranked link pairs $n_{YY}$ and $n_{NY}$ (for conflict appearance and resolution, respectively) for each of the methods LR (blue) and SB (cyan), and for their overlap.

## Methods

**Data collection.** During the academic year 2010–2011, we collected data on teamwork evaluation and preferences of 86 chemical engineering students that are grouped into teams facing an open ended project that lasts 9 months. We collected our data through an online survey that includes questions to evaluate different aspects of teamwork. We administered the same survey twice (December, survey I, and May, survey II).

Our sample consists of sixteen teams with the same structure: a fourth year student that plays the role of team leader and first year team members; the number of team members for which we have complete data (that is, that reported in both surveys I and II) ranges from 3 to 7, with most teams having 5–6 members (Table 1). Team membership was determined as follows. First, individuals were randomly assigned to one of four large groups. From each of these groups, four teams were defined so as to balance personality traits of their members (based on a personality test) but otherwise randomly.

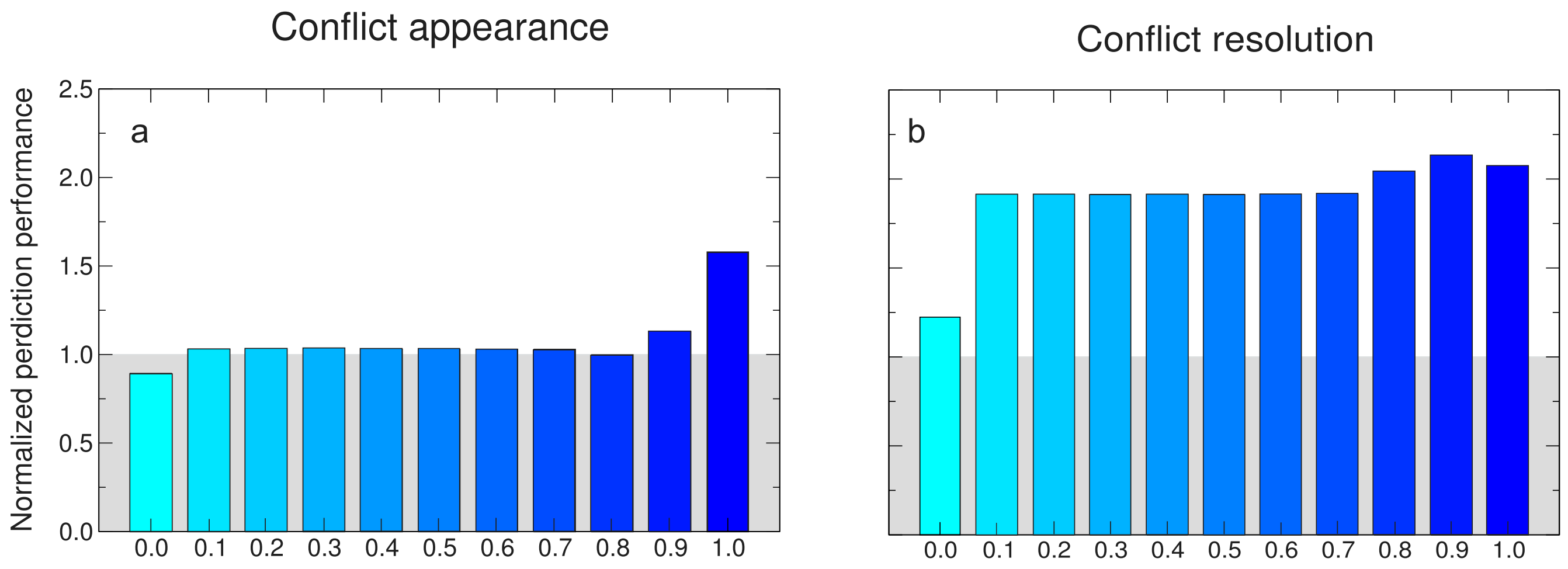

**Figure 3 | Hybrid scores for conflict prediction.** We introduce a hybrid score ($S^H$) obtained from the linear combination of the scores of both methods, $S^{LR}$ and $S^{SB}$ (Text and Methods). We plot the normalized prediction performance of the hybrid score for conflict appearance, (a), and conflict resolution, (b), as a function of a parameter $\alpha \in [0, 1]$ that enables us to interpolate between $S^H(\alpha = 0) = S^{SB}$, and $S^H(\alpha = 1) = S^{LR}$.

In our analysis, we focus in the answers of two yes/no questions from the survey: 1) "Would you choose this person to work with you in a new team?"; 2) "Would you choose this leader to lead a new team?". We use the answer to these questions as a proxy for the quality of interactions among team members. Thus, a yes answer would be a positive interaction whereas a no answer would be indicative of conflict among the pair of team members.

With this information we construct a directed network for each one of the surveys I and II, in which the link from member $A$ to member $B$ can be of two types $l^{I}_{AB} = Y$ or $l^{II}_{AB} = N$. We only consider interactions between pairs of team members for which we have complete information, that is, when both members have answered both surveys.

**Conflict prediction methods.** *Link reliability score***.** We have extended the formalism developed in[32] to obtain the link reliability score $S^{LR}$, that is, the probability that a link from member A to member B is of type $Y$ in survey II, $l^{II}_{AB} = Y$, given the observation $N^O$ of all interactions reported in survey I.

The fundamental assumption of this approach is that the structure of the network of interactions within a team can be satisfactorily accounted for by a model $M$, which is unknown but belongs to a family $\mathcal{M}$ of models. Then, the probability that a link from member A to member B is of type $Y$, $l_{AB} = Y$, given the observed network $N^O$ is[32]

$$S^{LR} = p\left(l^{II}_{AB} = Y|N^O\right) = \int_{\mathcal{M}} dM\, p\left(l^{II}_{AB} = Y|M\right) p\left(M|N^O\right), \tag{1}$$

To estimate this integral we rewrite it, using Bayes theorem, as[32,33]

$$S^{LR} = p\left(l^{II}_{AB} = Y|N^O\right) = \frac{\int_{\mathcal{M}} dM\, p\left(l^{II}_{AB} = Y|M\right) p\left(N^O|M\right) p(M)}{\int_{\mathcal{M}} dM\, p\left(N^O|M\right) p(M)}. \tag{2}$$

Here, $p(N^O|M)$ is the probability of the observed interactions given a model and $p(M)$ is the *a priori* probability of a model, which we assume to be model-independent $p(M) = \text{const}$.

For the family of stochastic block models, we have that for a given partition of team members into groups, there is a probability $Q(\alpha, \beta)$ of there being a link of type $Y$ from a member in group $\alpha$ to a member in group $\beta$, and a probability $(1 - Q(\alpha, \beta))$ of there being a link of type $N$. Note that because we are dealing with a directed network, $Q(\alpha, \beta)$ is not a symmetric matrix since for each block model team member A will be classified into two groups: a group for the outgoing links profile ($\sigma_{A\ \text{out}}$), and a group for the incoming links profile ($\sigma_{A\ \text{in}}$). Thus, if $A$ belongs group $\alpha$ for outgoing links and $B$ to group $\beta$ for incoming links, we have that[33]

$$p\left(l^{II}_{AB} = Y|M\right) = Q(\alpha,\beta); \tag{3}$$

and

$$p\left(N^O|M\right) = \prod_{\alpha \in G_{\text{out}}, \beta \in G_{\text{in}}} Q(\alpha,\beta)^{n^Y(\alpha,\beta)} \left(1 - Q(\alpha,\beta)^{n^N(\alpha,\beta)}\right), \tag{4}$$

where $n^{Y/N}(\alpha, \beta)$ is the number of links of type $Y/N$ between member groups $\alpha$ and $\beta$, and $G_{\text{out/in}}$ is the set of groups for outgoing/incoming link profiles in block model $M$. Additionally, the integral over all models in $\mathcal{M}$ can be separated into a sum over all possible partitions of the members into outgoing and incoming link groups, and an integral over all possible values of $Q(\alpha, \beta)$. These integrals can be carried out exactly to get[32,33]

$$S^{LR} = p\left(\left(l^{II}_{AB} = Y|N^O\right) = \frac{1}{Z} \sum_P \left(\frac{n^Y(\sigma_{A\ \text{out}}, \sigma_{B\ \text{in}}) + 1}{n(\sigma_{A\ \text{out}}, \sigma_{B\ \text{in}}) + 2}\right) \exp(-H(P)), \tag{5}$$

where the sum is over all partitions of the team members into outgoing and incoming link groups, $n(\sigma_{A\ \text{out}}, \sigma_{B\ \text{in}}) = \Sigma_{T:\{Y,N\}}\, n^T(\sigma_{A\ \text{out}}, \sigma_{B\ \text{in}})$ is the total number of known interactions from groups $\sigma_{A\ \text{out}}$ and $\sigma_{B\ \text{in}}$, and $H(P)$ is a function that depends on the partition only

$$H(P) = \sum_{\alpha,\beta} \left[\ln(n(\alpha,\beta) + 1)! - \sum_{T:\{Y,N\}} \ln\left(n^T(\alpha,\beta)\right)!\right]. \tag{6}$$

The sum in Eq. (5) can be estimated using the Metropolis algorithm to sample partitions[32].

*Structural balance score***.** To obtain $S^{SB}_{AB}$ we look at all the possible triads of members in a team that include members A and B. Then, we count the number of balanced triads $t_{\text{bal}}(Y)$ when $l_{AB} = Y$, and the number of balanced triads $t_{\text{bal}}(N)$ when $l_{AB} = N$. We then obtain

$$S^{SB}_{AB} = t_{\text{bal}}(Y) - t_{\text{bal}}(N). \tag{7}$$

According to structural balance theory, a balanced triad is one in which there is an odd number of positive reciprocal interactions. A positive reciprocal interaction is one such that $l_{AB} = l_{BA} = Y$.

For all the graphs and discussions in the main text we use the definition above for the structural balance-based score. One may argue, however, that this definition is somewhat restrictive because if an interaction is not reciprocal to start with, each of the nodes can only improve overall balance by switching, but never by staying in the same state. Therefore, we also consider here a second structural balance score $S^{SB2}_{AB}$

$$S^{SB2}_{AB} = S^{SB2}_{BA} = t_{\text{bal}}(l_{AB} = l_{BA} = Y) - t_{\text{bal}}(l_{AB} = l_{BA} = N). \tag{8}$$

that is, the difference between the number of balanced triangles when both links $AB$ and $BA$ are positive and the number of balanced triangles when both links $AB$ and $BA$ are negative. As we show in Figure S1 (Supplementary Information), this definition does not yield higher predictive power than the one discussed in the main text.

*Hybrid scores***.** For each link $AB$, we obtain a hybrid score $S^H_{AB}$ by combining $LR$ and $SB$ sores. However, because $S^{LR}$ is normalized, $S^{LR} \in [0, 1]$, and $S^{SB}$ is not, we first need to normalize $S^{SB}$. For each link $AB$ within team $T$, we obtain the normalized $SB$ score $\tilde{S}^{SB}_{AB}$ as follows

$$\tilde{S}^{SB}_{AB} = \frac{S^{SB}_{AB} - S^{SB}_{T\,\text{min}}}{S^{SB}_{T\,\text{max}} - S^{SB}_{T\,\text{min}}}, \tag{9}$$

where $S^{SB}_{T\,\text{min}}$ and $S^{SB}_{T\,\text{max}}$ are the minimum and maximum scores in team $T$, respectively.

We then obtain a hybrid score for each link $S^H_{AB}$ by linearly combining $S^{LR}_{AB}$ and $\tilde{S}^{SB}_{AB}$,

$$S^H_{AB} = \alpha S^{LR}_{AB} + (1-\alpha)\tilde{S}^{SB}_{AB}, \tag{10}$$

where $\alpha \in [0, 1]$ is the parameter that allows us to interpolate between $SB$ ($\alpha = 0$) and $LR$ ($\alpha = 1$) score rankings.

## Acknowledgements

We thank L.A.N. Amaral, A. Aguilar-Mogas, A. Gavaldà-Miralles, A. Godoy-Lorite, F.A. Massucci, T. Vallès-Català, and B. Uzzi for their comments and suggestions. This work was supported by a James S. McDonnell Foundation Research Award (RG and MSP), grants PIRG-GA-2010-277166 (RG) and PIRG-GA-2010-268342 (MSP) from the European Union, and grant FIS2010-18639 (RG and MSP) from the Spanish Ministerio de Economía y Competitividad. NRA acknowledges Universitat Rovira i Virgili for a Ph.D scholarhip.

## Author contributions

N.R.A., T.G., M.S.P. and R.G. designed the research protocol. N.R.A. and T.G. designed the survey. N.R.A. and R.G. administered surveys and collected data. N.R.A., M.S.P. and R.G. analyzed the data. N.R.A., M.S.P. and R.G. wrote the manuscript. N.R.A., T.G., R.G. and M.S.P. reviewed the manuscript.

## Additional information

**Supplementary information** accompanies this paper at http://www.nature.com/scientificreports

**Competing financial interests:** The authors declare no competing financial interests.